\documentclass[pra, showpacs, twocolumn, floatfix,superscriptaddress]{revtex4}
\usepackage{amsmath}
\usepackage{graphicx}
\usepackage{subfigure}
\usepackage{amsmath, amsfonts, amssymb, bm}

\allowdisplaybreaks[2]

\newcommand{\figtype}{-small}

\begin{document}
\title{Direct detection of $n$-particle atomic correlations via light scattering}
\author{Luling \surname{Jin}}
\affiliation{Max-Planck Institut f\"ur Kernphysik, Saupfercheckweg 1, D-69117 Heidelberg, Germany}
\affiliation{Department of Physics, Northwest University,  Xi'an 710069, Shaanxi, China}

\author{Mihai \surname{Macovei}}
\affiliation{Max-Planck Institut f\"ur Kernphysik, Saupfercheckweg 1, D-69117 Heidelberg, Germany}

\author{J\"{o}rg \surname{Evers}}
\affiliation{Max-Planck Institut f\"ur Kernphysik, Saupfercheckweg 1, D-69117 Heidelberg, Germany}

\date{\today}
\begin{abstract}
The creation and direct detection of $n$-particle atomic correlations  in ensembles of atoms is  investigated. For this, we study an ensemble of laser-driven atoms in which either a dipole-dipole or a Rydberg-Rydberg interaction leads to the formation of  correlations between the internal degrees of freedom of the atom. We show that light scattering can be used to imprint information about these correlations onto light, and reveal how this information can be extracted from the statistical properties of the scattered light. As main result we find that observation in certain detection directions allows to directly and individually measure  $n$-particle atomic correlations. Complementary, we discuss a method to experimentally determine the interesting detection positions.
\end{abstract}



\maketitle

\section{Introduction}
The study of correlated quantum systems is ubiquitous in many branches of physics.
In particular cold atoms offer various implementations of few- and many-body quantum systems with tunable strong long-range interaction~\cite{coldgas}, such as polar molecules~\cite{polar} or Rydberg atoms~\cite{ryd1,ryd2}. 
A systematic way of characterizing these systems is via suitable correlation functions.  Naturally, in particular higher-order correlations are of interest, since they generally indicate more complex dynamics, and since their knowledge is required to completely characterize a quantum system. 
Unfortunately, however, higher-order correlations are generally hard to observe. 

Different methods have been suggested to measure correlations, depending on the size and structure of the system, and on the type of correlations. In larger systems, correlations in position or momentum space are typically detected via interferometric approaches~\cite{coldgas}.
An alternative approach is to coherently transfer correlations to light, e.g., via off-resonant light scattering. This on the one hand can be used for imaging cold gases~\cite{2dmott}, but also has been suggested for the measurement of spin correlations~\cite{transfer1,transfer2,transfer3,transfer4,2dmott} or atom density correlations~\cite{density}. Enhancement of the measurement signal has been proposed via cavities~\cite{cavity} and Bragg scattering~\cite{bragg}.

Similar imaging challenges arise in smaller systems. In terms of light scattering, in particular the conceptionally simple case of few atoms trapped in periodic potentials has received much attention. 
Next to setups with uncorrelated atoms~\cite{uncorr1,uncorr2,uncorr3,uncorr4}, also systems with interacting atoms have been considered. It has been shown that the interaction between the particles leads to modifications of the scattered light~\cite{corr1,corr2,corr3,corr4} which can for example be used to measure distance of the scattering particles~\cite{distance}. 

This interest in interacting few-particle systems has recently been revived in the context of Rydberg atoms, which provide a tunable long-range distance. A setup with two optically trapped atoms has been used to  demonstrate the Rydberg blockade~\cite{2dipole}, to entangle two atoms~\cite{entang}, and to implement a two-atom gate~\cite{gate}. Extensions to more atoms are highly desirable. For example,  it has been suggested that the addition of a third atom within the blockade radius could lead to a significant modification of the blockade property~\cite{anti}. Similarly, the entangling and gate operations naturally scale to systems with more than two atoms. The characterization of such systems requires methods to determine correlations between the internal electronic degrees of freedom of few trapped particles.

\begin{figure}[b]
\includegraphics[width=7cm]{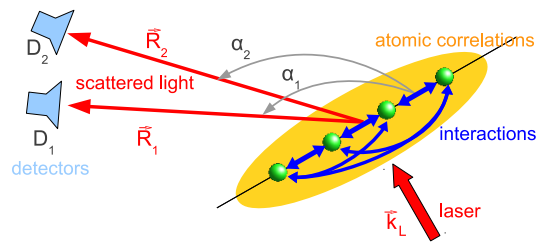}
\caption{\label{fig1}(Color online)  A chain of $N$ atoms  driven by a resonant laser field. Detectors $D_{1}$ and $D_{2}$ measure the second order correlation function of the scattered photons. The atoms interact, which leads to the formation of $n$-particle atomic correlations. Key objective is the direct measurement of these few-particle atomic correlations via the scattered light.}
\end{figure}

Motivated by this, here, we study the creation and detection of higher-order atomic correlations in an ensemble of atoms. We aim at detecting  correlations between the internal degrees of freedom of the atoms generated by interactions. For this, we analyze a chain of laser-driven atoms, in which  the atomic correlations arise only due to the interaction between the atoms, see Fig.~\ref{fig1}.  Our focus is on the direct detection of atomic correlations by detecting the intensity-intensity correlation function $G^{(2)}$ of the light scattered by the atoms, with light scattering as the simplest transfer of atomic properties onto the light properties. We show that $G^{(2)}$ contains contributions which can be traced back to $2$-, $3$- and $4$-particle atomic correlations. A closer analysis reveals that observation in particular detection directions allows to detect the different $n$-particle atomic correlations individually. The method is independent on the coupling generating the correlations, and results are given both for the case of dipole-dipole interaction (DDI) and Rydberg-Rydberg interaction (RRI).
%

\section{Theoretical considerations}
\subsection{Model}
We start by investigating a system of $N$ atoms distributed in a linear chain, as shown in Fig.~\ref{fig1}. To include both DDI and RRI, the identical particles are modelled as three-level atoms in ladder configuration with ground state $|g\rangle$,  excited state $|e\rangle$, and  Rydberg state $|r\rangle$~\cite{weidemuller,adams}, and are located at positions $\vec{r}_i$ ($i\in \{1,2, \cdots, N\}$), with separation  $r_{ij} = |\vec{r}_i - \vec{r}_j|$. The lower [upper] transitions are driven by resonant laser fields with Rabi frequency $\Omega_p$ [$\Omega_c$], which propagate perpendicular to the atom chain. 
 In the electric dipole, Born-Markov and rotating wave approximations and in a suitable interaction picture, the system's master equation is~\cite{agarwal,fs}
\begin{align}
\label{master}
\frac{\partial \rho}{\partial t} &= \frac{1}{i\hbar} [V, \rho ] 
 - \sum_{i=1}^{N}\frac{\gamma_c}{2}( [A_{re}^{(i)},A_{er}^{(i)}\rho]  + \textrm{H.c.}) \nonumber \\
& -\sum_{i,j=1}^N
\frac{\gamma^{(ij)}_p}{2}( [A_{eg}^{(i)},A_{ge}^{(j)}\rho] + \textrm{H.c.}) \,,
\end{align}
with $V=V_L + V_{dd}+V_{RR}$. 
Here $A_{\alpha\beta}^{(i)}=|\alpha\rangle_i\langle \beta|$ is an operator of atom $i$, and  $\gamma_p \equiv \gamma^{(ii)}_p$ [$\gamma_c$] is the spontaneous decay rate on the lower [upper] transition.
The atom-laser interaction is given by 
\begin{align}
V_L = \hbar  \sum_{i} (\Omega_p A_{eg}^{(i)}  + \Omega_c A_{re}^{(i)} + \textrm{H.c.})\,.
\end{align}

For the DDI case, we reduce the system to the two lower states by setting $\Omega_c=0$, and by dropping $V_{RR}$.  The coherent part of the DDI is given by
\begin{align}
V_{dd} = - \hbar \sum_{i\neq j}    \Omega^{(ij)} A_{eg}^{(i)}A_{ge}^{(j)}\,.
\end{align}
The parameters $\gamma^{(ij)}_p$ and $\Omega^{(ij)}$ ($i\neq j$) are the usual  DDI coupling constants characterized by the tensor~\cite{fs}
\begin{align}
\chi_{pq}(\vec{r}) &= \frac{k_0^3}{4\pi\epsilon_0}\left \{\delta_{pq}\left (\frac1\eta + \frac{i}{\eta^2} - \frac{1}{\eta^3} \right ) \right . \nonumber\\
&\left . - \frac{[\vec{r}]_p\,[\vec{r}]_q}{r^2} \left (\frac{1}{\eta} +\frac{3i}{\eta^2} -\frac{3}{\eta^3}\right ) \right \}e^{i\eta}\,,
\end{align}
where $\eta = |\vec{k}_L|\cdot |\vec{r}|$. From this tensor, the coupling constants are obtained as
\begin{subequations}
\begin{align}
\gamma^{(ij)}_p &= \frac{1}{\hbar}(\vec{d}^T \textrm{Im}[\chi(\vec{r}_{ij})] \vec{d}^*)\,,\\
\Omega^{(ij)} &= \frac{1}{\hbar}(\vec{d}^T \textrm{Re}[\chi(\vec{r}_{ij})] \vec{d}^*)\,,
\end{align}
\end{subequations}
with $\vec{d}$ the involved dipole moments which we assume to be perpendicular to the interatomic distance vectors.  

For the RRI case,  we assume large distance on the scale of the lower transition wavelength, such that the DDI on the lower transition vanishes, $\gamma^{(ij)}_p=0=\Omega^{(ij)}$ for $i\neq j$. The RRI coupling is described by
\begin{align}
V_{RR} = \hbar \sum_{i\neq j}   V_{ij}A_{rr}^{(i)}A_{rr}^{(j)}\,,
\end{align}
and the coupling constant is taken as $V_{ij} = C_6 / |r_{ij}|^6$.

\subsection{Observables}
From the steady state solution of Eq.~(\ref{master}), the scattered light intensity and the second order correlation function can be calculated as~\cite{glauber,scullybook}
\begin{subequations}
\label{Corr}
\begin{align}
G^{(1)}&\propto  \sum_{i,j}\langle 
A_{eg}^{(i)}A_{ge}^{(j)} \rangle e^{i\vec k_{1}\cdot\vec r_{ij}}\,, \\
G^{(2)} &\propto \sum_{i,j,k,l}\langle 
A_{eg}^{(i)}A_{eg}^{(j)}A_{ge}^{(k)}A_{ge}^{(l)}  
\rangle e^{i\cdot \vec k_{1}\vec r_{il}+i  \vec k_{2}\cdot \vec r_{jk}}\,, 
\end{align}
\end{subequations}
up to prefactors which we neglect in the following. 
%
%
We next define correlation functions $U^{(1)}$ and $U^{(2)}$ excluding the correlations between different atoms as
\begin{subequations}
\begin{align}
U^{(1)}&\propto  \sum_{i,j}\langle 
A_{eg}^{(i)}A_{ge}^{(j)} \rangle_U e^{i\vec k_{1}\cdot\vec r_{ij}}\,, \\
U^{(2)} &\propto \sum_{i,j,k,l}\langle 
A_{eg}^{(i)}A_{eg}^{(j)}A_{ge}^{(k)}A_{ge}^{(l)}  
\rangle_U e^{i\cdot \vec k_{1}\vec r_{il}+i  \vec k_{2}\cdot \vec r_{jk}}\,, 
\end{align}
\end{subequations}
%
These are obtained from 
$G^{(1)}$ and $G^{(2)}$ by the replacement $\langle\cdot\rangle \to  \langle\cdot\rangle_U$. Here, $\langle\cdot\rangle_U$ is the correlation function including correlations of individual atoms, but explicitly excluding correlations between  different atoms. For example, for any  operators $A_i$ and $B_i$ operating on atom $i$ and any operator $C_j$ operating on a different atom $j\neq i$, 
\begin{align}
\langle A_i B_i C_j\rangle_U=\langle A_i B_i \rangle\langle C_j\rangle\,.
\end{align}
We analyze these correlations as a function of the detector positions. For this, we relate the outgoing wave vectors $\vec{k}_1$ and $\vec{k}_2$ to the detector positions. If, for example, the arrangement is as in Fig.~\ref{fig1} with a linear chain driven perpendicularly to the chain axis, then ($n\in\{1,2\}$)
\begin{align}
  \vec{k}_n\cdot\vec{r}_{ij} = (2\pi/\lambda)\, r_{ij}\, \cos \alpha_n\,,
\end{align}
and the detection directions $\alpha_n$ immediately translate to detection positions $\vec{R}_n$ in the far field. With this convention, in the following, we will interchangably use detector positions as arguments such as in $G^{(1)}(\vec R)$ and $G^{(2)}(\vec{R}_1, \vec{R}_2)$ and the detection angles $\alpha_n$.

Finally, we define the normalized second order correlation function~\cite{glauber,scullybook}
\begin{align}
g^{(2)}(\vec R_1, \vec R_2) = \frac{G^{(2)}(\vec R_1, \vec R_2)}{G^{(1)}(\vec R_1)\:G^{(1)}(\vec R_2)}\,,
\end{align}
which is a measure for the probability of detecting scattered photons at $\vec R_1$ and $\vec R_2$ in coincidence, and provides information about the photon statistics. $g^{(2)}=1$ indicates Poissonian photon statistics, whereas $g^{(2)}<1$ and $g^{(2)}>1$ correspond to sub-Poissonian and super-Poissonian photon statistics, respectively.

%
\begin{figure}[t]
\centering
\includegraphics[width=8cm]{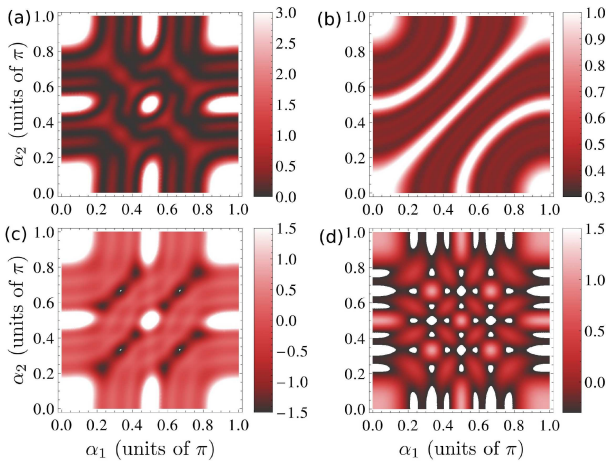}
\caption{\label{fig-G}(Color online) Second order correlation function $G^{(2)}$ and its contributions for dipole-dipole interacting atoms. (a) shows the full second-order function $G^{(2)}$, and (b) shows the contribution $G_2$, (c) $G_3$ and (d) $G_4$. The results are plotted scaled by $10^6$ and against the positions of the two detectors $\alpha_1$ and $\alpha_2$ for $N=4$, $\Omega_p = 0.01 \gamma_p$ 
and $r_{i,i+1}=\lambda_p$.}
\end{figure}


\section{Results}
\subsection{Without interaction}
We first analyze the case without interaction between the atoms, $V_{ij}=\gamma^{(ij)}_p=\Omega^{(ij)}=0$ for $i\neq j$, which will help in interpreting our later results. It is important to distinguish between correlations in the light and correlations between the atoms. The former are characterized by the correlation functions in Eq.~(\ref{Corr}), whereas the latter are characterized by the difference of $G$ and $U$. As expected, we found that without interaction, $G^{(1)} = U^{(1)}$ and $G^{(2)} = U^{(2)}$, such that no correlations between the atoms are created without the interaction. Nevertheless, the photonic correlation functions have contributions involving two, three or four particles at the same time, which is in analogy to the interference pattern from a multi slit aperture. These contributions, however, do not indicate atomic correlations. Without interaction, we found that
\begin{align}
\label{int-simple}
G^{(1)}(\vec R)=U^{(1)}(\vec R) = I\,N + C\sum_{i\neq j} e^{i\vec{k_1}\cdot\vec{r}_{ik}}\,,
\end{align}
where $I=\langle A_{ee}^{(i)}\rangle$ as a measure for the incoherent scattered light intensity and $C=|\langle A_{eg}^{(i)}\rangle|^2$ as a measure for the coherent intensity are the same for all atoms $i$. The second-order correlation function can be decomposed as 
\begin{align}
G^{(2)} = U^{(2)} = G_{2} + G_{3} + G_{4}\,,
\end{align}
in which $G_n$ contains contributions involving $n$ atoms, i.e., those in which the indices $i,j,k,l$ in Eq.~(\ref{Corr}) take on $n$ different values. This decomposition allows for an interpretation of the different contributions.
First, there are no contributions $G_1$ with $i=j=k=l$ arising from single atoms, which  cannot emit two photons at the same time.
$G_{2}$ are contributions involving two different atoms, related to the Hanbury-Brown-Twiss type interference~\cite{scullybook}. 
%
%
$G_{3}$ consists of three-particle contributions.  These occur even though the detectors only register two photons in each event, because  $G_{3}$ combines first- and second-order interference, as indicated by the fact that this term is proportional to $IC$.
Finally, $G_{4}$ is a contribution involving four different atoms. This is the highest-order contribution to the second-order correlation function, since it involves only four photon operators. It is proportional to $C^2$, relying fully on first- and second-order interference.


\begin{figure}[t]
\centering
\includegraphics[width=7cm]{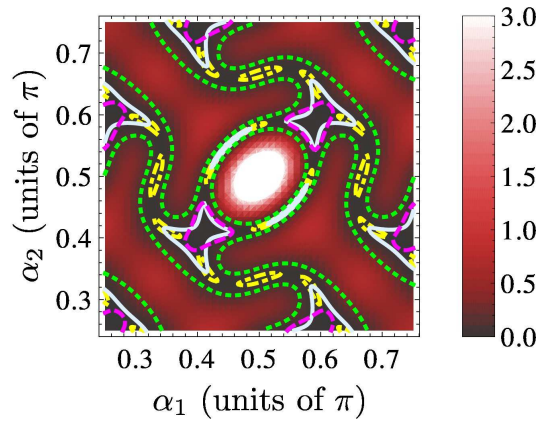}
\caption{\label{fig-pos}(Color online) Detection positions in which $n$-atom correlations can directly be measured for dipole-dipole interacting atoms. The dotted green contour indicates $\mathcal{C}_2=0$,
and the dashed magenta contour $\mathcal{C}_4=0$. In these directions, $G^{(2)}$ is solely due to 2-atom or 4-atom correlations, respectively. The solid light blue contour shows $\mathcal{C}_3/ G^{(2)}= 10$, and the dot-dashed yellow contour $\mathcal{C}_4/ G^{(2)}= 10$. In these areas 3-atom or 4-atom correlations strongly reduce $G^{(2)}$, respectively.  The background shows $G^{(2)}$ as in Fig.~\ref{fig-G}(a).}
\end{figure}

\subsection{With interaction}

 We now turn to our main interest, the creation and detection of $n$-particle atomic correlations. We found that these appear as soon as the atoms interact. Then, $G^{(1)} \neq U^{(1)}$ and $G^{(2)} \neq U^{(2)}$, and the different atoms in general assume unequal steady states. Therefore no simple expression as in Eq.~(\ref{int-simple}) can be obtained. Nevertheless, we found that the decomposition into $n$-particle contributions $G^{(2)} = G_{2} + G_{3} + G_{4}$ and analogously $U^{(2)} = U_{2} + U_{3} + U_{4}$ still persists, allowing for a convenient identification of $n$-particle atomic correlations.
We in particular ask whether it is possible to identify and directly verify the presence of 2-, 3- and 4-particle atomic correlations. For this, we define
\begin{align}
\mathcal{C}_n = G^{(2)} - G_n + U_n\,, \label{contour}
\end{align}
which is the full second-order correlation function $G^{(2)}$ with the contribution due to $n$-particle atomic correlations subtracted. If we now find detector positions for which $\mathcal{C}_n = 0$, then $G^{(2)} = G_n - U_n$. Therefore, at these detection position, the measured value for $G^{(2)}$ can directly be identified as the contribution arising from the $n$-atom correlations, such that the contribution of $n$-atom correlations can be detected and quantified individually.
On the other hand, if  $\mathcal{C}_n \gg G^{(2)}$, then $n$-atom correlations decrease the full $G^{(2)}$. This criterion again indicates atomic correlations.  But in general it does not guarantee an individual measurement of a single $n$-atom correlations. For example, strong negative contributions from two different $n$-atom correlations together could lead to a vanishing value of $G^{(2)}$.

\begin{figure}[t]
\includegraphics[width=0.9\columnwidth]{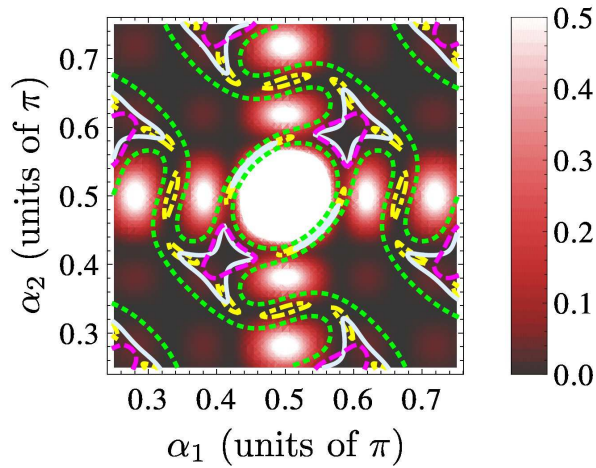}
\caption{\label{fig-int}(Color online) Emitted light intensity for dipole-dipole interacting atoms as a function of detection positions $\alpha_1$ and $\alpha_2$. The figure shows the product $G^{(1)}(\vec R_1)\,G^{(1)}(\vec R_2)$ and is thus nonzero only if both detectors register light. Parameters and contours are as in Fig.~\ref{fig-pos}. }
\end{figure}
\begin{figure}[t]
\centering
\includegraphics[width=0.9\columnwidth]{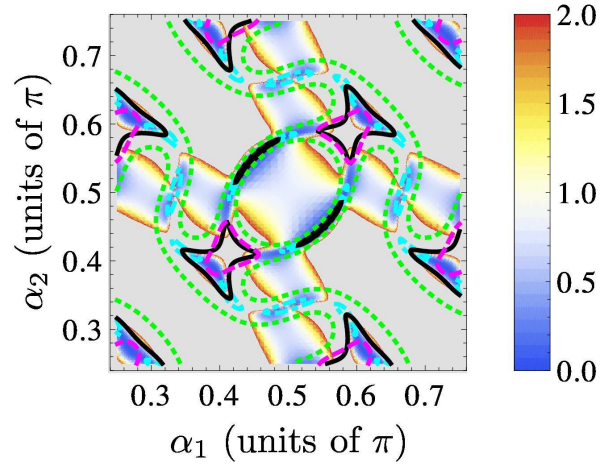}
\caption{\label{fig-g2}(Color online) Photon statistics $g^{(2)}(\tau=0)$ for dipole-dipole interacting atoms as a function of detection positions $\alpha_1$ and $\alpha_2$. Contours and parameters are as in Fig.~\ref{fig-pos}.}
\end{figure}

\subsection{Dipole-dipole interaction}
We start with results for DDI. In Fig.~\ref{fig-G}, results for the second-order correlation function $G^{(2)}$ and its individual contributions are shown. The main structure as a function of the detection angles is due to the spatial dependence on the detection positions encoded in the exponential functions in Eq.~(\ref{Corr}). In (a) it can be seen that around detector positions $\alpha_i \in \{0,\pi/2,\pi\}$ the full correlation function assumes large values, which can be attributed to the geometrical structure of the atom arrangement. Such interference peaks can be used to deduce the precise positioning of the scattering particles in an experiment~\cite{2dmott}, but do not provide further insight in the correlations between the particles. Around these maxima a rather complicated pattern occurs, which will turn out to contain information on the atomic correlations. Interestingly, the individual $n$-particle contributions in (b-d) have quite different dependences on the detection positions. Also, we find that the 3- and 4-particle contributions can be large in magnitude and both positive or negative, such that cancellations in $G^{(2)}$ occur.

\begin{figure}[t]
\includegraphics[width=0.9\columnwidth]{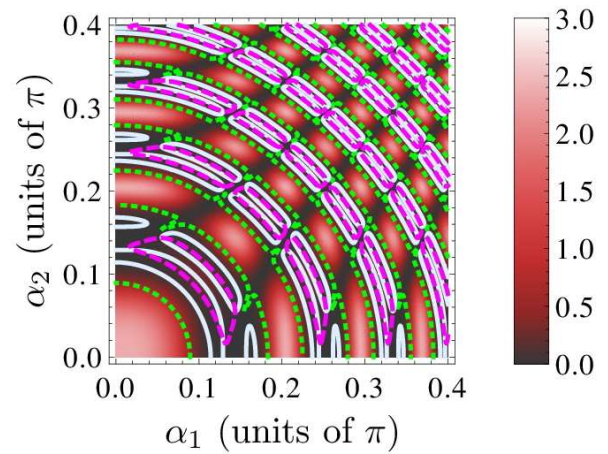}
\caption{\label{fig-ryd}(Color online) Detection positions in which $n$-atom correlations can directly be measured for Rydberg-Rydberg interacting atoms. The dotted green contour indicates $\mathcal{C}_2=0$, and the dashed magenta contour $\mathcal{C}_4=0$. In these directions, $G^{(2)}$ is solely due to 2-atom or 4-atom correlations, respectively. The solid light blue contour shows $\mathcal{C}_3/ G^{(2)}= 5$, such that $3$-atom correlations dominate.  The background shows $G^{(2)}$. Parameters are $\Omega_p = 0.01 \gamma_p$, $\Omega_c =  \gamma_p$, $\gamma_c = 0.05 \gamma_p$, $C_6  = 2\pi\times 50$GHz~$\mu$m$^6$, and $r_{i,i+1}=5\lambda_p$.}
\end{figure}

\begin{figure}[t]
\centering
\includegraphics[width=0.9\columnwidth]{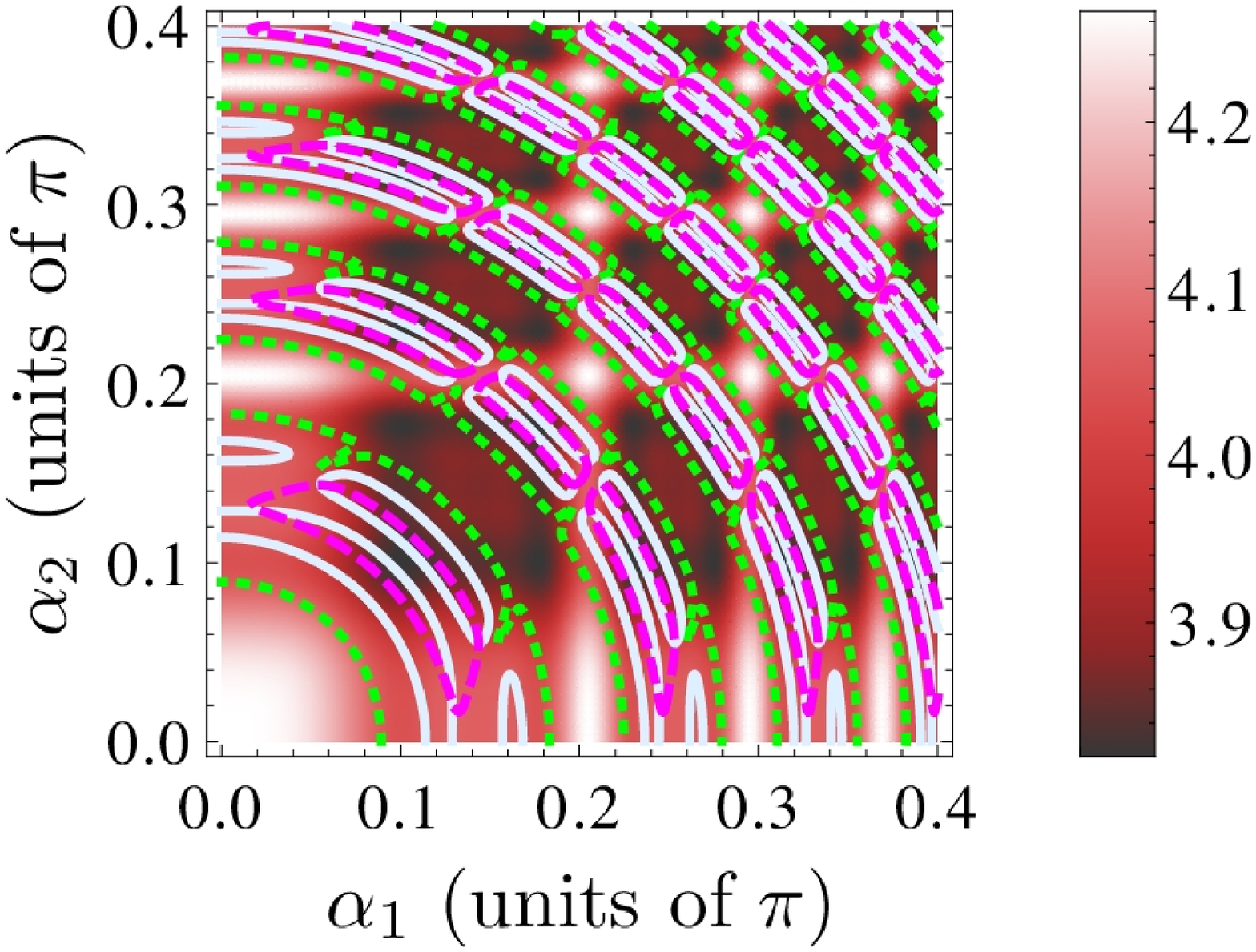}
\caption{\label{fig-intr}(Color online) Emitted light intensity for Rydberg-Rydberg interacting atoms as a function of detection positions $\alpha_1$ and $\alpha_2$. The figure shows the product $G^{(1)}(\vec R_1)\,G^{(1)}(\vec R_2)$ and is thus nonzero only if both detectors register light. Parameters and contours are as in Fig.~\ref{fig-ryd}. 
 }
\end{figure}

\begin{figure}[t]
\centering
\includegraphics[width=0.9\columnwidth]{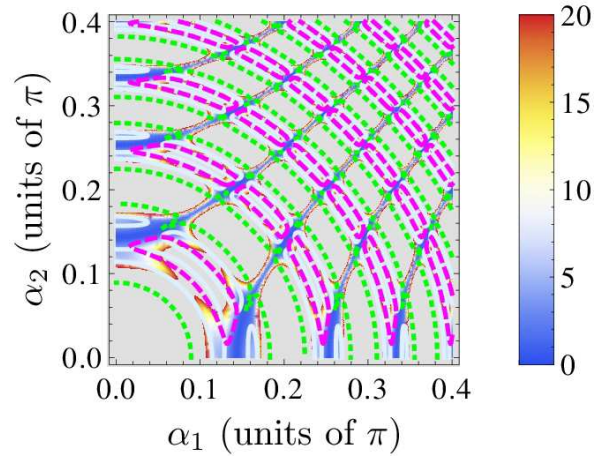}
\caption{\label{fig-g2rr}(Color online) Photon statistics $g^{(2)}(\tau=0)$ for Rydberg-Rydberg interacting atoms as a function of detection positions $\alpha_1$ and $\alpha_2$. Contours and parameters are as in Fig.~\ref{fig-ryd}.}
\end{figure}

Fig.~\ref{fig-pos} shows results on the direct and individual detection of $n$-particle correlations. The contours plotted using Eq.~(\ref{contour}) indicate that there are detection ranges in which $G^{(2)}$ is non-zero only due to $2$-particle or $4$-particle atomic correlations. At these positions, without the respective $n$-particle atomic correlation, the value of $G^{(2)}$ would be zero, and any detection signal directly can be traced back to the atomic correlations. Note that such regions do not occur for 3-particle atomic correlations. On the other hand, there are detection directions in which $\mathcal{C}_3/ G^{(2)}\geq 10$ or $\mathcal{C}_4/ G^{(2)}\geq 10$. In these areas, the 3- or 4-particle atomic correlations dominate the value of $G^{(2)}$. 

Fig.~\ref{fig-int} shows the corresponding emitted light intensity as a function of the detection positions $\alpha_1$ and $\alpha_2$. More specifically, the figure shows the product $G^{(1)}(\vec R_1)\,G^{(1)}(\vec R_2)$ and is thus nonzero only if both detectors register light. As in Fig.~\ref{fig-pos}, the contours indicate detection positions in which $n$-particle atomic correlations can be directly measured.  It can be seen that all contours overlap with regions of non-zero intensity registered by the detectors. Parameters are as in Fig.~\ref{fig-pos}.

Finally, Fig.~\ref{fig-g2} shows the photon statistics $g^{(2)}(\tau=0)$ in dependence on the detection positions $\alpha_1$ and $\alpha_2$. The range of values is from $0$ to $2$. Areas with values larger that $2$ are shaded in light gray. As in Fig.~\ref{fig-pos}, the contours indicate detection positions in which $n$-particle atomic correlations can be directly measured, and the parameters are chosen the same.  It can be seen that depending on the detection direction, light of all statistics from sub-Poissonian to super-Poissonian is emitted. This complicated pattern of the photon statistics, however, in general is unrelated to the  correlations between the atoms, and in particular also appears for uncorrelated atoms. Around the positions of the maximum intensity  at $\alpha_1 \approx \alpha_2 \approx \pi/2$, $g^{(2)}\approx 1$, consistent with coherent light required for full constructive interference. Interestingly, already each of the contour lines indicating detection areas in which certain $n$-atom correlations can be measured span a range of photon statistics including sub- and super-Poissonian statistics. Along each contour, the type of correlation between the atoms remains the same, while the photon statistic imprinted on the scattered photons varies with the detection directions.

\subsection{Rydberg-Rydberg interaction}
 Next, we show that the method to determine $n$-particle atomic correlations also works with RRI. We found that the results are qualitatively comparable to DDI, and the interpretations are the same.  However, Fig.~\ref{fig-ryd} shows that $G^{(2)}$ has a very different structure as for the DDI case. This in part is because in the RRI case, the interparticle distance is larger than in the DDI case, resulting in a different overall interference pattern. But also, the RRI has a very different character than the DDI, which is reflected in the correlation pattern. From Fig.~\ref{fig-ryd}, detection regions can be identified in which $C_i=0$  for $i\in\{2,4\}$, such that $2$- and $4$-particle correlations can be measured directly and individually. Also, regions of large $C_3/ G^{(2)} \gg5$ exist, providing access to the $3$-particle correlations.

Fig.~\ref{fig-intr} shows the emitted light intensity for Rydberg-Rybderg interacting atoms as a function of the detection positions $\alpha_1$ and $\alpha_2$, with parameters as in Fig.~\ref{fig-ryd}. As in Fig.~\ref{fig-int} for the DDI case, the figure shows the product $G^{(1)}(\vec R_1)\,G^{(1)}(\vec R_2)$ and is thus nonzero only if both detectors register light. As in Fig.~\ref{fig-ryd}, the contours indicate detection positions in which $n$-particle atomic correlations can be directly measured for the RRI case.  As in the DDI case, it can be seen that all contours overlap with regions of non-zero intensity registered by the detectors.

The photon statistics $g^{(2)}(\tau=0)$ in dependence on the detection positions $\alpha_1$ and $\alpha_2$ for the RRI case is shown in Fig.~\ref{fig-g2rr}.  Areas with values larger than $20$ are shaded in light gray. In most detection directions, $g^{(2)}(\tau=0)>1$ is observed. However, the directions in which 2-particle correlations can be observed also reach into regions with $g^{(2)}(\tau=0)<1$. Parameters are as in Fig.~\ref{fig-ryd}.

\section{Discussion and Summary}

\begin{figure}[t]
\includegraphics[width=0.95\columnwidth]{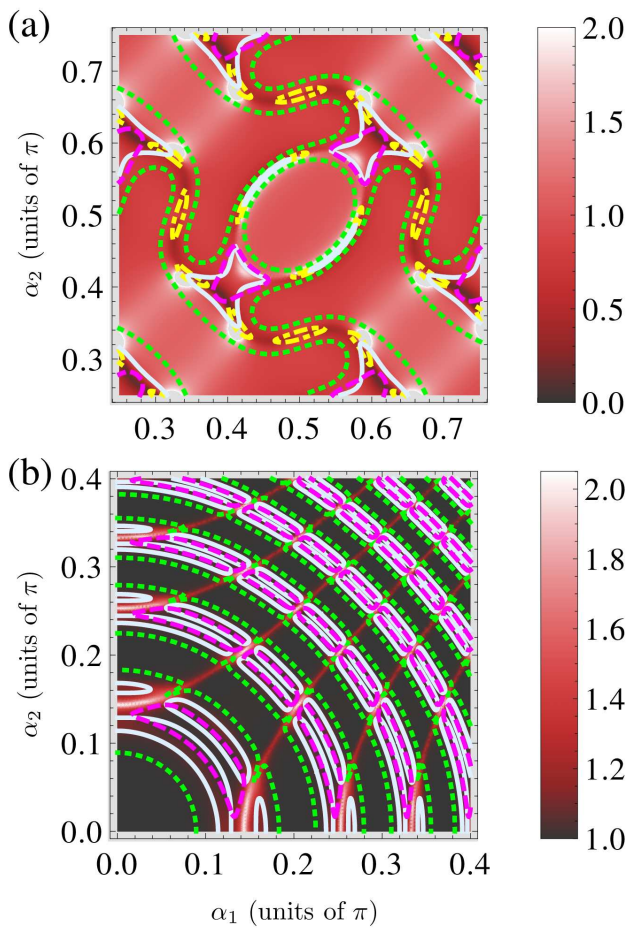}
\caption{\label{fig-exp}(Color online) Method to experimentally determination detection directions to identify $n$-particle correlations. (a) Dipole-dipole interacting atoms. The ratio of $G^{(2)}$ for  interatomic spacing $\lambda_p$ to $G^{(2)}$ for interatomic spacing $1.5\lambda_p$ with scaled coordinates is shown. (b) Rydberg-Rydberg interaction. The ratio of $G^{(2)}$ with $\Omega_c = 0.01 \gamma_p$ to the case with $\Omega_c = 0.05 \gamma_p$ is shown. Overlays show the theoretically predicted detection directions.}
\end{figure}

Our calculations rely on a knowledge of the positions of the atoms. But in an actual experiment, the atom positions may not be known precisely enough. To overcome this problem, there are different options to measure the atom positions. In our setup, $G^{(2)}$ needs to be measured, such that automatically also $G^{(1)}$ becomes available. The required atom position information is contained in $G^{(1)}$. The configuration of atom positions leads to a pattern of interference maxima in the position-dependent scattered light intensity, which can be used to determine the atom configuration. This works particularly well if the atoms are arranged in a periodic pattern, such as in an optical lattice. This is in essence the same method as it is used in crystallography to reveal the spatial arrangement of a crystal from scattered light. If a periodic potential is used, then the scattered light still allows to detect ``defects'' like empty sites of the lattice. 
On the other hand, alternative imaging methods have been developed to image the positions of atoms in an optical lattice~\cite{2dmott}. 
Finally, if a long-range interaction like that between Rydberg states generates the atomic correlations, then the atoms can be placed at distances which allow for an individual imaging. Thereby, again the position information can be gained.

A more interesting option indeed is to modify the multiatom correlations. In the following, we discuss several methods to do so. First, if the atoms are placed in an optical lattice, then by changing the angle between the lattice beams, the periodicity of the lattice can be changed. Of course, this will change the scattered light pattern considerably. But if the atoms are uncorrelated, then  the atoms evolve individually, such that the expectation values of atomic operators $\langle \cdot \rangle$ entering the expression Eq.~(\ref{Corr}) for $G^{(2)}$ do not depend on the positions of the atoms. Then, the position dependence of $G^{(2)}$ only arises from the exponential factors in Eq.~(\ref{Corr}). For example, for two lattice spacings $s_1$ and $s_2$ and a suitable alignment of the lattice with respect to the incident field, one finds  phases $\vec k_{n}\vec r_{il} = (2\pi/\lambda) s_1 \cos(\alpha_n)$ and $\vec k_{n}\vec r_{il} = (2\pi/\lambda) s_2 \cos(\bar{\alpha}_n)$ for the two lattice spacings in the exponents of Eq.~(\ref{Corr}) ($n=1,2$). By equating $s_1 \cos \alpha_n = s_2 \cos \bar{\alpha}_n$, a transformation between the two detection angles $\alpha_n$ and $\bar{\alpha}_n$  can be obtained. Applying this transformation  to the results with lattice spacing $s_2$, the results for lattice spacing $s_1$ are obtained. We verified this by calculating the scattered light for two different lattice spacings without coupling between the atoms, i.e., without correlations. When we applied the scaling transformation to one set of data, we obtained exactly the data for the other distance. 
This scaling, however, fails if distant-dependent interactions and thus correlations between the atoms are present. Then, the atomic expectation values entering Eq.~(\ref{Corr}) do depend on distance. The scaled data obtained with lattice spacing $s_2$ will therefore differ from the results of lattice spacing $s_1$ in those detection directions, in which correlations exist.
Thus next, we repeated the procedure with the interaction between the atoms switched on. Now, the difference between the first data set and the scaled second data set is non-zero. 
An example is shown in Fig.~\ref{fig-exp}(a). It can be seen that the differing  parts reproduce the detection directions we predict in order to measure the atomic correlations to a good degree.  This is not surprising, as the scaling is expected to fail exactly at those detection directions at which correlations modify the result. An experiment could thus measure the scattered light for two lattice spacings, and determine the positions in which correlations can be detected by searching for differences between the scaled first data set and the second data set.   Note, however, that the presence of correlations does not necessarily imply a difference between two measurements at different lattice spacings. Therefore, there is no one-to-one correspondence between the theoretical contours and the positions determined using a measurement as outlined above. 
A different option is to make use of the angular dependence of the coupling in 1-d atomic arrangements.  If the dipole moments are aligned at different angles with respect to the trap axis, then also the coupling changes~\cite{dir}.

For RRI atoms, it is even simpler to obtain two measurements to compare with in order to determine potential directions in which correlations can be measured. We found that it is sufficient to take the ratio of two measurements with different $\Omega_c$, without changing the atom positions or the interaction strength. As Fig.~\ref{fig-exp}(b) shows, again the ratio of the two measurements exhibit structures which resemble the detection directions predicted theoretically. Next to the experimental determination of detection directions, this allows for a direct comparison of theory and experiment.

It can be expected that higher-order correlation functions $G^{(m)}$ with $m>2$ enable one to detect $n$-atom correlations with $n>4$ in a straightforward extension of the present work. Increasing the number of atoms, however, while still measuring $G^{(2)}$, would not provide direct access to higher order correlations, since only correlations between 4 atomic operators enter the expression of $G^{(2)}$. Nevertheless, measurements with larger structures are desirable, as a measurement of the scattered light in particular detection directions is required, such that count rates become important. 
An alternative could be to combine our method  with strong laser driving in modified reservoirs~\cite{melk,me} to increase photon scattering. 

In this work, we have presented results on regular structures of atoms, which could be realized, e.g., by placing atoms in an optical lattice. The regular structure essentially leads to the formation of interference maxima in particular directions of the light scattering, in which constructive interference of the paths via the different atoms appears. Our results on the correlation measurements, however, are not associated with these interference maxima arising due to the regular structure of the atoms. Therefore we expect that our method also works with random positions of the atoms. To substantiate this expectation, as an example, we have randomly modified the interparticle distances in the chain to $r_{12}=1.3\lambda$, $r_{23}=0.6\lambda$, $r_{34}=0.4\lambda$ in the DDI case, and calculated the results correponding to Fig.~\ref{fig-pos}. Also in this case, contours could be found in which the different correlations can be measured individually.

Concluding, we have demonstrated a method to directly and individually measure $n$-particle atomic correlations by scattering light off of a small ensemble of atoms. 
In our model, the correlations are generated by the interaction between the atoms, and we have demonstrated results for both the dipole-dipole and the Rydberg-Rydberg interaction case. In a broader context, our results shed light on the interaction-induced formation of correlations and in particular enable an order-by-order comparison of theory and experiment,  and thus open perspectives for studying the correlated dynamics of ensembles of interacting atoms.

\begin{acknowledgements}
LLJ  acknowledges hospitality at the Max Planck Institut f\"ur Kernphysik and the support from NFSC (Grant No. 61108006).
\end{acknowledgements}


\end{document}